%% file: main.tex
\documentclass[10pt,conference]{IEEEtran}

\usepackage{cite}
\usepackage{amsmath,amssymb,amsfonts}
\usepackage{graphicx}
\usepackage{eso-pic}
\usepackage{flafter}
\usepackage{booktabs}
\usepackage{tabularx}
\usepackage{xcolor}
\usepackage{listings}
\usepackage{colortbl,pifont}
\usepackage{url}
\usepackage{mdframed}

\definecolor{tablehead}{HTML}{E8EEF4}
\definecolor{tableband}{HTML}{F4F6F8}
\definecolor{tablesummary}{HTML}{E6F1EE}
\newcommand{\rowpadtop}{\rule{0pt}{\dimexpr\ht\strutbox+0.65ex\relax}}
\newcommand{\rowpadbot}{\rule[-\dimexpr\dp\strutbox+0.4ex\relax]{0pt}{0pt}}
\newcommand{\cmidrulefill}[1]{\noalign{{\color{#1}\hrule height\cmidrulewidth}\vskip-\cmidrulewidth}}

\newcommand{\modellogo}[1]{\raisebox{-0.14em}{\includegraphics[height=0.9em]{#1.png}}\hspace{0.35em}}

\providecommand{\Description}[1]{}

\newcommand{\benchname}{\textsc{FeatureArc}}

\newcommand{\rqanswer}[2]{%
  \begin{mdframed}[backgroundcolor=black!7,hidealllines=true,
    innerleftmargin=\fboxsep,innerrightmargin=\fboxsep,
    innertopmargin=\fboxsep,innerbottommargin=\fboxsep,
    skipabove=\smallskipamount,skipbelow=\smallskipamount]
    \noindent\textbf{Answer to #1:} #2
  \end{mdframed}}

\begin{document}

\AddToShipoutPictureFG*{%
  \AtPageUpperLeft{%
    \put(\LenToUnit{\dimexpr1in+\oddsidemargin\relax},-\LenToUnit{42pt}){%
      \includegraphics[height=24pt]{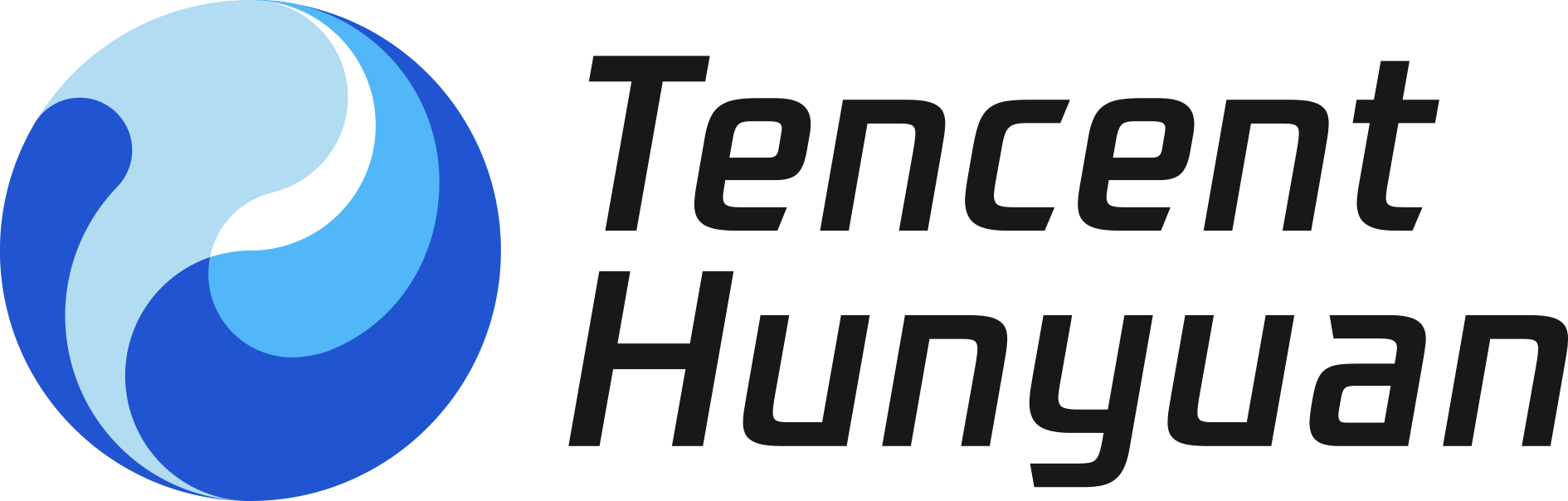}}%
    \put(\LenToUnit{\dimexpr1in+\oddsidemargin\relax},-\LenToUnit{47pt}){%
      {\color{black!25}\rule{\textwidth}{0.4pt}}}%
  }%
}

\title{\benchname: Benchmarking Coding Agents on Evolving Features}

\author{
\IEEEauthorblockN{Yunlong Wang\textsuperscript{1},
Alyssa Wu\textsuperscript{1},
Cuiyun Gao\textsuperscript{2},
Jia Feng\textsuperscript{1},
Songyuanyi Lu\textsuperscript{1},
Zhichao Hu\textsuperscript{1}}
\IEEEauthorblockA{\textsuperscript{1}Hunyuan Team, Tencent\\
\textsuperscript{2}The Chinese University of Hong Kong\\
\{joeyolwang, alyssawwu, plusfeng, alinasyylu, frankiehu\}@tencent.com, cygao@cse.cuhk.edu.hk\\
\textsuperscript{*}Corresponding authors: Alyssa Wu and Cuiyun Gao.}
}

\maketitle

\begin{abstract}
A software feature usually evolves over time: developers extend its first version as new use cases and requirements arrive, and each extension builds on code written earlier.
Most feature benchmarks evaluate a single request on a fixed codebase. Feature evolution requires evaluating successive extensions on an agent's own earlier code and checking whether the feature remains correct.
We introduce \benchname{} to evaluate coding agents on evolving features, together with a pipeline that constructs its tasks from repository histories.
The benchmark combines dependencies between successive stages of the same feature, explicit requirements and interface contracts, and executable checks across four programming languages.
The pipeline validates fail-to-pass (F2P) tests for new behavior and pass-to-pass (P2P) tests for existing behavior on the reference code before and after each stage.
We evaluated 10 coding agents on 73 manually reviewed tasks of 2 to 10 stages.
Each agent works in one persistent session and workspace without external test feedback or rollback.
The evaluation reveals a gap between first-stage and whole-feature success: pooled success falls from 62.7\% at the first stage to 28.5\% when the whole feature is tested on the final code.
Trajectory analysis identifies inherited defects as a common failure mechanism, present in 72.9\% of failed later stages.
These earlier errors often remain undetected: in stages following a failure, agents leave the earlier error unnoticed about 95\% of the time, and their self-tests rarely revisit earlier stages.
Evolving a feature autonomously, without human checks between stages, therefore remains a substantial challenge for current coding agents.
At Tencent, the pipeline extends internal evaluations to this setting and guides the development of Hunyuan, Tencent's family of large language models.
\end{abstract}


\input{01-introduction}
\input{02-motivation}
\input{03-approach}

\input{04-study-design}
\input{05-results}
\input{06-lessons}

\input{08-related-work}
\input{09-conclusion}

\clearpage

\end{document}

%% file: 01-introduction.tex
\section{Introduction}
\label{sec:introduction}

\input{fig1-delegation}

Software features evolve as developers extend existing code to meet new requirements~\cite{lehman1980programs,rajlich2000staged}.
We describe this evolution as a sequence of \emph{stages}, from the initial implementation to successive extensions.
Coding agents now inspect repositories, edit code, and run tests autonomously~\cite{yang2024sweagent,wang2025openhands}.
On SWE-bench, which asks agents to resolve individual GitHub issues~\cite{jimenez2024swebench}, the best model resolved 1.96\% at release in 2023; agents now resolve more than 70\% of its 500-issue, human-validated Verified subset~\cite{openai2024swebenchverified,swebench2026leaderboard}.
In 37\% of the 13,000 real sessions collected in SWE-chat, agents wrote virtually all committed code~\cite{baumann2026swechat}.
Developers have also begun delegating successive extensions of a whole feature without intervening checks.
Each extension then inherits the agent's earlier code, including errors the agent has not noticed.
We ask: \emph{can an agent that implements single requirements well keep a feature correct while extending its own code as requirements evolve?}

\input{benchmark-comparison}

Evaluating this ability requires agents to sustain a feature while checking their own earlier work.
User feedback between stages determines which implementation and verification responsibilities remain with the agent.
Similar to the levels of driving automation~\cite{sae2021j3016} and of AI-agent autonomy~\cite{feng2025levels}, we distinguish four levels of delegation for \emph{feature implementation}: at \textbf{L1} the user tells the agent how to fix a failed stage, at \textbf{L2} the user points out the failing behavior, at \textbf{L3} the user only says whether a stage passed, and at \textbf{L4} the user only gives the next specification and says nothing about earlier stages (Figure~\ref{fig:delegation}).
L4 feature implementation requires the agent to find earlier errors itself, whereas at L1--L3 the user reports failed stages.
Tencent's internal multi-stage feature evaluations covered only L1--L3, creating a practical need for L4 evaluation of sustained development.
An earlier attempt to build L4 tasks by splitting a single feature request into stages produced weak dependencies: the resulting stages behaved largely as independent single-stage tasks, limiting their ability to test successive extensions.
L4 therefore needs stages that depend on each other by construction.
We recover these stages from repository histories to capture how developers actually extended a feature to meet new requirements, with each extension building on earlier implementations.

Existing feature benchmarks address parts of this setting, but leave a gap in evaluating same-feature evolution from repository histories at L4.
As shown in Table~\ref{tab:benchmark-comparison}, most pose a single feature request on a fixed starting state~\cite{li2025feabench,chen2025featbench,zhou2026featurebench}, so there is no later stage for an error to affect.
Benchmarks with several stages evaluate different development settings: CCBench~\cite{codecrafters2026ccbench} gives all stages in one instruction; SWE-STEPS~\cite{shastry2026swesteps} returns test feedback after each stage and chains pull requests that need not belong to one feature; and SlopCodeBench~\cite{orlanski2026slopcodebench}, EvoCode-Bench~\cite{shen2026evocode}, and EvoWeb~\cite{nukrai2026evoweb} use stages written by hand or generated from similar products, which do not follow how a feature actually grew in a repository.
Among the benchmarks compared in Table~\ref{tab:benchmark-comparison}, \benchname{} is the only one that combines historical same-feature dependencies, sequential requests, and continued evaluation on the agent's own code without hidden-test feedback.
Ticks, triangles, and crosses indicate full, partial, and absent support; automated construction covers tasks, environments, and tests, with human review allowed.

We address this gap with \benchname{} and an automated pipeline for constructing its tasks.
The benchmark has three complementary properties.
First, \textbf{historical same-feature dependencies} require agents to extend an existing implementation: the pipeline recovers successive stages of one user-visible feature from repository histories.
Second, \textbf{requirements-grounded verification} makes each extension explicit and testable: a requirements document and interface contract specify the expected behavior, while fail-to-pass (F2P) tests and pass-to-pass (P2P) regression tests~\cite{yoo2012regression} check new and existing behavior against reference states.
A final check evaluates the complete feature on the agent's resulting code.
Third, \textbf{diverse repository contexts} expose agents to feature evolution in Python, TypeScript, Rust, and Go, across seven application domains.
We select the evaluation set through stratified sampling, multi-model trajectory screening for task defects, and manual checks (Section~\ref{sec:eval-set}).
At L4, each agent receives specifications sequentially in a persistent session and workspace; hidden-test results are withheld, and the agent carries its own code forward after every stage.

The results expose three weaknesses of the evaluated coding agents on evolving-feature tasks.
First, \textbf{whole-feature success is less than half of first-stage success}: pooled success falls from 62.7\% at the first stage to 28.5\% on the whole feature, and even the strongest agent, Claude Opus~5.5~\cite{anthropic2026opus55}, completes only 49.3\% of the features, although it passes 83.6\% of first stages.
Second, \textbf{inherited errors often persist}: after a failed stage, agents pass the next stage's F2P tests in only 22.6\% of cases, versus 70.2\% after a passed one, whereas breaking earlier behavior affects only 3.3\% of features whose new behavior works.
Third, \textbf{agents rarely notice their own earlier failures}: in 94.9\% of the stages that follow a failed stage, the agent never notices the error.
Evolving a feature without human checks between stages therefore remains a substantial challenge for current coding agents.

At Tencent, we use this pipeline to add L4 feature implementation to internal evaluations and assess Hunyuan~4 Preview, a pre-release model in the Hunyuan family~\cite{tencent2026hunyuan}, alongside external agents.
The evaluation distinguishes implementing a new request from detecting errors in the code on which it depends.
This distinction gives the Hunyuan team concrete evidence about failures that first-stage scores do not reveal, as discussed in Section~\ref{sec:tencent}.

Our contributions are:
\begin{itemize}
  \item \textbf{A benchmark and construction pipeline} that combine historical same-feature dependencies, requirements-grounded verification, and diverse repository contexts for L4 feature implementation.
  \item \textbf{An empirical study of coding agents} on \benchname{} that quantifies the gap between first-stage and whole-feature success and documents the persistence of inherited errors and limitations in agents' self-verification.
  \item \textbf{An industrial application at Tencent}: extending internal evaluation to L4 and diagnosing Hunyuan's development needs through feature-level outcomes and trajectory evidence.
\end{itemize}

%% file: fig1-delegation.tex
\begin{figure}[t]
  \centering
  \includegraphics[width=\columnwidth]{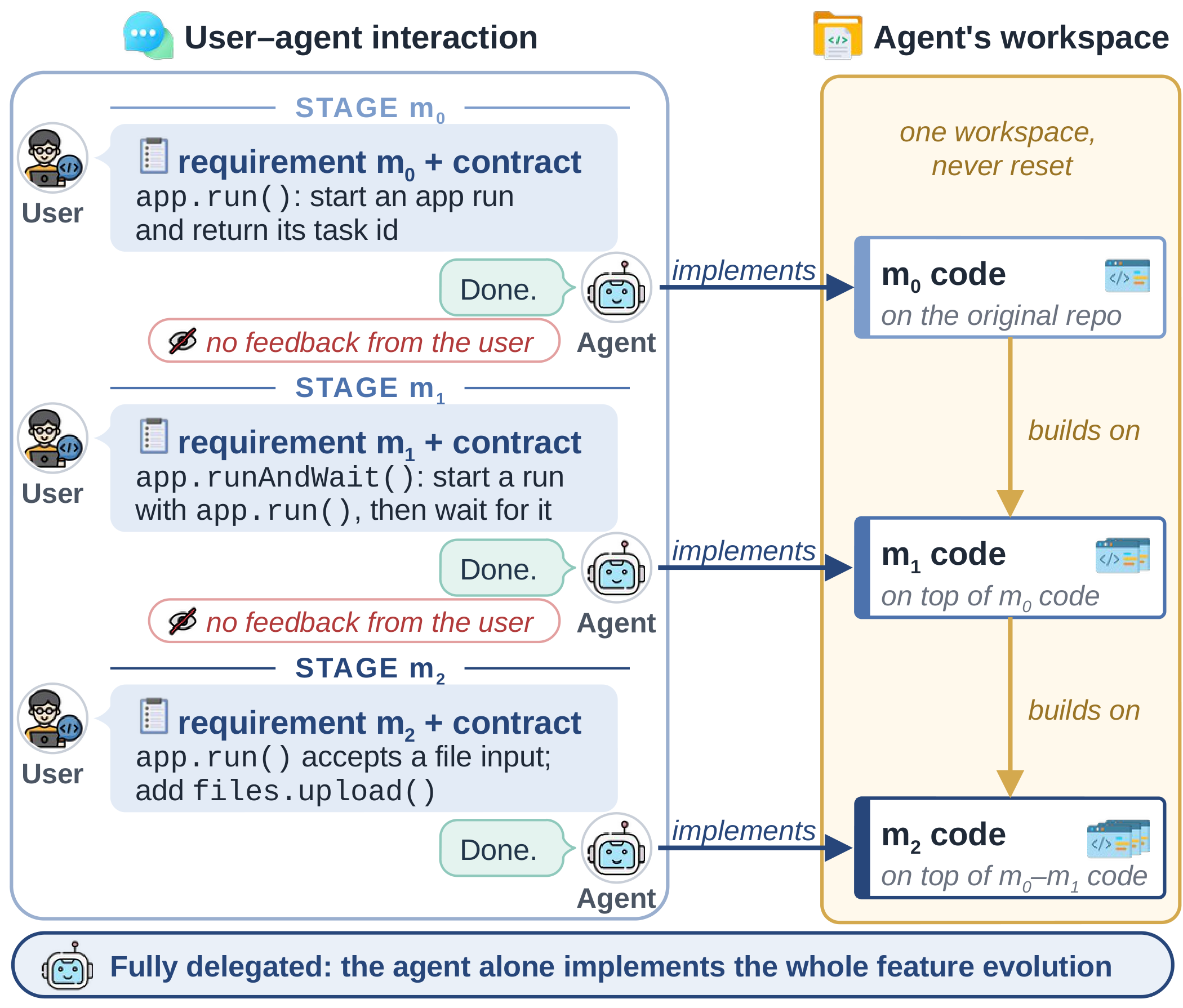}
  \caption{\textbf{L4 feature implementation.}}
  \Description{Left, user-agent interaction over three stages. At each stage the user gives a specification
  (requirements and contract): m0, app.run() starts an app run and returns its task id; m1, app.runAndWait() starts a run
  with app.run() and waits for it; m2, app.run() accepts a file input and files.upload() is added. The
  agent replies Done, and the user gives no feedback. Right, the agent's workspace, one workspace that is
  never reset: m0 code on the original repository, m1 code on top of m0 code, and m2 code on top of m0 and
  m1 code. Footer: fully delegated, the agent alone implements the whole feature evolution.}
  \label{fig:delegation}
\end{figure}

%% file: benchmark-comparison.tex
\begin{table}[!t]
  \centering
  \footnotesize
  \setlength{\tabcolsep}{1.5pt}
  \renewcommand{\arraystretch}{1.20}
  \setlength{\aboverulesep}{0pt}
  \setlength{\belowrulesep}{0pt}
  \definecolor{cmpnavy}{RGB}{35,61,88}
  \definecolor{cmpblue}{RGB}{35,82,134}
  \definecolor{cmpteal}{RGB}{15,106,99}
  \definecolor{cmpheader}{RGB}{231,238,246}
  \definecolor{cmpsection}{RGB}{242,246,250}
  \definecolor{cmpstripe}{RGB}{248,250,252}
  \definecolor{cmpours}{RGB}{221,239,236}
  \newcommand{\cmphead}[1]{\textcolor{cmpnavy}{\textbf{#1}}}
  \newcommand{\cmpstackh}[2]{\scriptsize\begin{tabular}[c]{@{}c@{}}\textcolor{cmpnavy}{\textbf{#1}}\\[-1.5pt]\textcolor{cmpnavy}{\textbf{#2}}\end{tabular}}
  \newcommand{\cmpstackt}[3]{\scriptsize\begin{tabular}[c]{@{}c@{}}\textcolor{cmpnavy}{\textbf{#1}}\\[-1.5pt]\textcolor{cmpnavy}{\textbf{#2}}\\[-1.5pt]\textcolor{cmpnavy}{\textbf{#3}}\end{tabular}}
  \newcommand{\cmpstackc}[2]{\begin{tabular}[c]{@{}c@{}}\textcolor{cmpnavy}{\textbf{#1}}\\[-1.5pt]\textcolor{cmpnavy}{\textbf{#2}}\end{tabular}}
  \newcommand{\cmpgroup}[1]{\rowcolor{cmpsection}\multicolumn{6}{l}{\rule{0pt}{2.3ex}\rule[-0.8ex]{0pt}{0pt}\textcolor{cmpblue}{\textbf{#1}}}\\}
  \newcommand{\cmpyes}{\textcolor{cmpteal}{\ding{51}}}
  \newcommand{\cmpno}{\textcolor[RGB]{170,63,57}{\ding{55}}}
  \newcommand{\cmppart}{\textcolor[RGB]{174,113,28}{\ensuremath{\triangle}}}
  \caption{Comparison of feature benchmarks.}
  \label{tab:benchmark-comparison}
  \begin{tabularx}{\columnwidth}{@{\hspace{2pt}}>{\raggedright\arraybackslash}X c c c c c@{\hspace{2pt}}}
    \toprule[0.9pt]
    \rowcolor{cmpheader}
    \rule{0pt}{3.0ex}\rule[-1.6ex]{0pt}{0pt}\cmphead{Benchmark}
      & \cmpstackh{Iteration on}{own code}
      & \cmpstackh{Single}{feature}
      & \cmpstackh{Repository}{history}
      & \cmpstackh{Automated}{construction}
      & \cmpstackh{Multiple}{languages}\\
    \midrule[0.5pt]
    \cmpgroup{One feature request per task}
    FEA-Bench~\cite{li2025feabench}
      & \cmpno & \cmpno & \cmpyes & \cmpyes & \cmpno\\
    \rowcolor{cmpstripe}
    FeatBench~\cite{chen2025featbench}
      & \cmpno & \cmpno & \cmpyes & \cmpyes & \cmpno\\
    FeatureBench~\cite{zhou2026featurebench}
      & \cmpno & \cmpno & \cmppart & \cmpyes & \cmpno\\
    \midrule[0.35pt]
    \cmpgroup{Several stages per task}
    CCBench~\cite{codecrafters2026ccbench}
      & \cmpno & \cmpyes & \cmppart & \cmpno & \cmpyes\\
    \rowcolor{cmpstripe}
    SWE-STEPS~\cite{shastry2026swesteps}
      & \cmpyes & \cmpno & \cmpyes & \cmpyes & \cmpno\\
    SlopCodeBench~\cite{orlanski2026slopcodebench}
      & \cmpyes & \cmppart & \cmpno & \cmpno & \cmpno\\
    \rowcolor{cmpstripe}
    EvoCode-Bench~\cite{shen2026evocode}
      & \cmpyes & \cmppart & \cmpno & \cmpno & \cmpyes\\
    EvoWeb~\cite{nukrai2026evoweb}
      & \cmpyes & \cmppart & \cmpno & \cmpyes & \cmpno\\
    \midrule[0.5pt]
    \rowcolor{cmpours}
    \rule{0pt}{2.4ex}\rule[-0.9ex]{0pt}{0pt}\textcolor{cmpteal}{\textbf{\benchname{} (ours)}}
      & \cmpyes & \cmpyes & \cmpyes & \cmpyes
      & \cmpyes\\
    \bottomrule[0.9pt]
  \end{tabularx}
\end{table}

%% file: 02-motivation.tex
\section{Task Formulation}
\label{sec:formulation}

We define a task as a sequence of stages and specify how the agent is evaluated under the L4 setting introduced in Section~\ref{sec:introduction}.

\subsection{Task Definition}
\label{sec:task-definition}

A \emph{task} follows one user-visible feature through an ordered sequence of $n$ stages, $m_0,\dots,m_{n-1}$, in one repository.
The first stage implements the initial version, and each later stage extends the same feature.
We call this sequence a \emph{same-feature chain}.

In the task illustrated in Figure~\ref{fig:delegation}, the first stage, $m_0$, adds \texttt{app.run()}, which starts a run and returns its task identifier.
The second stage, $m_1$, adds \texttt{app.runAndWait()}, which calls \texttt{app.run()} and waits for completion.
The third stage, $m_2$, extends the feature to accept file input and adds \texttt{files.upload()}.
Each stage builds on the agent's implementation of the preceding requirements.

The task provides an initial repository state $S_0$, a container image with the required dependencies, and a specification for each stage.
Each specification has two parts: a \emph{requirements document} describing the requested behavior and acceptance criteria, and an \emph{interface contract} stating the names, signatures, constants, and error messages needed to use that behavior.
The agent receives the specifications one at a time and implements each new request on top of its own earlier work.

Hidden \emph{fail-to-pass} (F2P) tests check the new behavior requested at a stage, and \emph{pass-to-pass} (P2P) tests check existing behavior~\cite{yoo2012regression}.
A separate pair of chain-level suites checks the complete feature on the final workspace.
Section~\ref{sec:construction} explains how we recover the stages, write their specifications, and validate these tests against reference code.

\subsection{Evaluation Process}
\label{sec:protocol}

\input{fig-2b-eval-protocol}
Figure~\ref{fig:protocol} shows the L4 evaluation process.
One session and workspace persist through the whole task.
The agent can inspect its code and run its own checks; reference patches, hidden test suites, and future specifications stay private.
When the agent finishes a stage, a hidden acceptance check, called a \textbf{stage gate}, runs its F2P and P2P suites.
The stage is the implementation request; the gate checks its result, and both suites must pass for the stage to pass.
\textbf{Failed stages are neither rolled back nor reported to the agent}: the next stage starts from the code the agent just produced.
After $m_{n-1}$, a \textbf{chain gate} runs the chain-level F2P and P2P suites on the final workspace to check the complete feature.

%% file: fig-2b-eval-protocol.tex
\begin{figure}[t]
  \centering
  \includegraphics[width=\columnwidth]{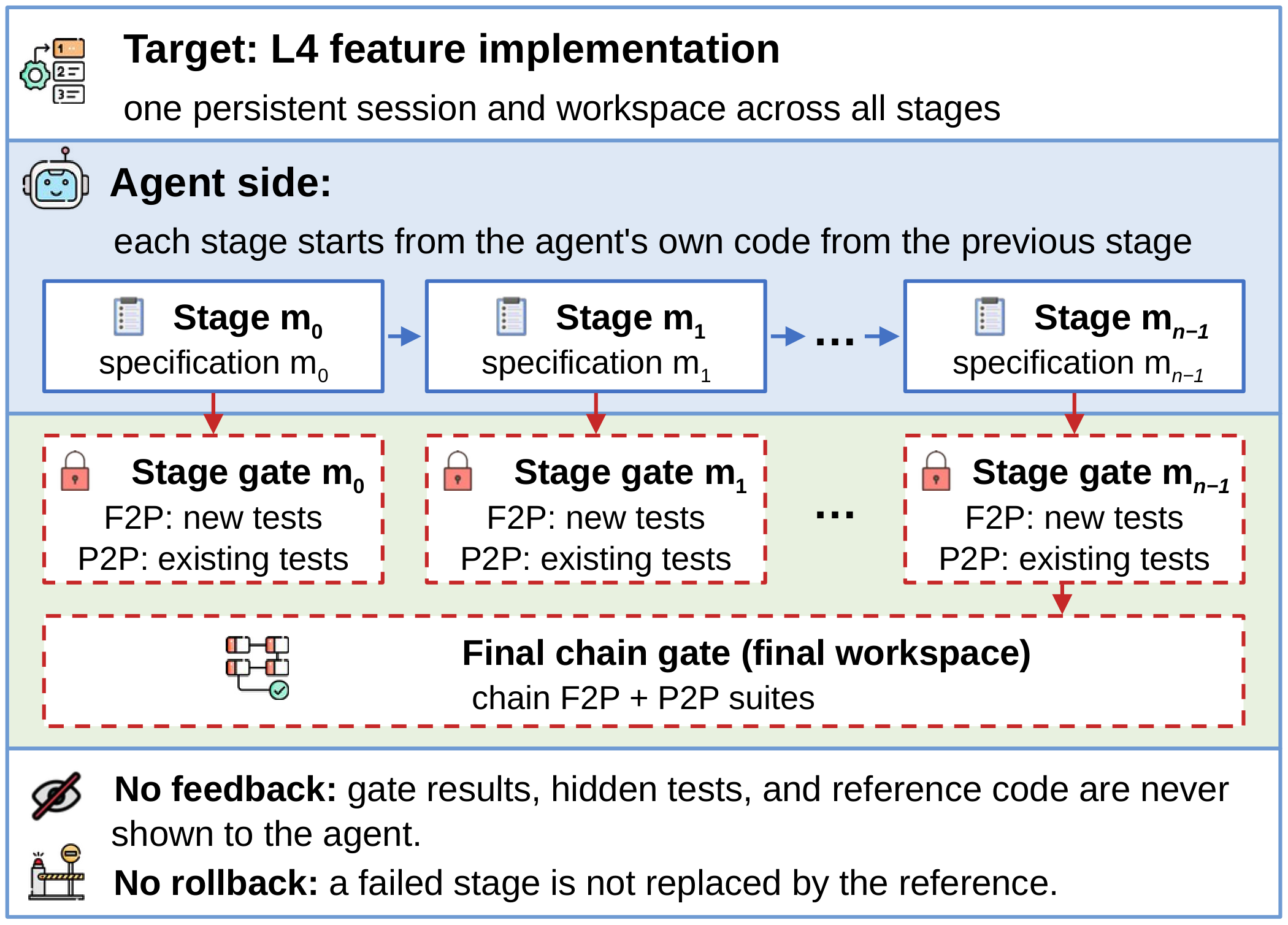}
  \caption{\textbf{L4 evaluation process.}}
  \Description{One persistent session and workspace across all stages. Agent side: stages m0, m1, through m(n-1),
  each with its specification, each starting from the agent's own code from the previous stage. Below each stage, a hidden stage gate runs F2P new tests and P2P
  existing tests; after m(n-1), a final chain gate runs the chain F2P and P2P suites on the final workspace.
  Footer: no feedback, since gate results, hidden tests, and reference code are never shown to the
  agent, and no rollback, since a failed stage is not replaced by the reference.}
  \label{fig:protocol}
\end{figure}

%% file: 03-approach.tex
\section{Benchmark Construction}
\label{sec:construction}

The construction pipeline turns repository histories into tasks that require an agent to extend the same feature on its own code.
Reliable evaluation requires dependent stages, explicit behavior specifications, and executable tests that distinguish the requested new behavior from existing behavior.
As shown in Figure~\ref{fig:construction-overview}, the pipeline addresses these requirements in five steps: repository collection and history analysis, evolution-chain mining, reference-state construction, environment and test validation, and specification generation.
We then select and manually review the evaluation set.

\input{fig-1-overview}

\subsection{Step 1: Repository Collection and History Analysis}
\label{sec:repo-selection}

To recover meaningful feature evolution, we need histories that contain successive extensions and repositories that can support executable evaluation.
As shown in Step~1 of Figure~\ref{fig:construction-overview}, we select repositories with at least 30 stars, permissive licenses, offline Linux builds, and code written mainly in Python, TypeScript, Rust, or Go.
The star threshold provides a basic indication of community interest; the other conditions support redistribution and repeatable execution.
Repository histories provide the feature changes and reference implementations used to construct and validate each task.
During evaluation, the agent extends its own earlier code, so each new requirement must be implemented against the workspace produced by its preceding edits.

Following a feature through file renames and module moves requires evidence beyond file names: changes to shared files suggest a relationship~\cite{zimmermann2005mining}, which we examine through the module structure and the intent of the associated commits.
We therefore build a module map from imports at the starting checkpoint and update it as we process commits in order.
The topology and semantic agents in Figure~\ref{fig:construction-overview} check complementary evidence: the former checks module structure, and the latter identifies commit intent and associated tests.
Their output supplies the historical relationships used in the next step.

\subsection{Step 2: Evolution-Chain Mining}
\label{sec:chain-mining}

To evaluate evolution of one feature, we must distinguish its successive extensions from unrelated changes that happen nearby in time or in the same files.
We represent candidate relationships in an evidence graph, illustrated in Step~2 of Figure~\ref{fig:construction-overview}.
Nodes are commits or small commit groups, and edges connect changes through shared code elements, file history, references, or imports.
This representation allows the pipeline to follow related changes even when other development occurs between them.

The four searches shown in the figure explore complementary parts of this graph: global beam search follows promising paths, anchor-guided search starts from identified feature changes, endpoint bridging connects related endpoints, and community-guided search examines groups of related changes.
An LLM review agent checks the proposed chains against source code, tests, and diffs.
A chain is retained only when every stage extends the same user-visible feature with observable, testable behavior.
The review also adds missed extensions, removes unrelated work, and adjusts stage boundaries.
The output is the mined same-feature chain used to construct the task.

\subsection{Step 3: Reference-State Construction}
\label{sec:reference-states}

To validate the contribution of each stage, we construct paired reference states that differ by the selected extension while retaining the surrounding repository context.
As shown in Step~3 of Figure~\ref{fig:construction-overview}, we label historical file patches by their role in the feature, its tests, and necessary dependencies, and assign the selected changes to stage patches.
We remove these patches from the final checkpoint to obtain the initial state $S_0$, then reapply them in order to obtain $S_1,\dots,S_n$.
Unselected changes remain as shared context.
These paired states let the next step test the contribution of each stage directly.

\subsection{Step 4: Environment and Test Validation}
\label{sec:test-validation}

To attribute a failure to the requested behavior, the task must run consistently and its checks must distinguish the code before and after the extension.
The environment agent in Step~4 of Figure~\ref{fig:construction-overview} prepares one offline container image for the chain and repairs it until the reference states build and their tests can be discovered.
The test-extraction agent packages historical tests with their required fixtures and helpers, and selects tests according to their observed outcomes on the paired states.
F2P tests must fail before the stage and pass afterward; P2P tests must pass on both states.
The chain-level suites undergo the same validation between $S_0$ and $S_n$, enabling the final check to cover the complete feature across all stages.

\subsection{Step 5: Specification Generation}
\label{sec:spec-generation}

To make the expected behavior accessible without revealing its implementation, each stage needs a behavioral description and an explicit interface contract.
As shown in Step~5 of Figure~\ref{fig:construction-overview}, a specification agent uses the stage patch and F2P tests to write the requirements document and contract.
The former states behavior and acceptance criteria; the latter specifies the names, signatures, constants, and error messages needed to use that behavior.
Before publication, the pipeline removes registered agent-generated documents and restores documents identified as revealing later implementations to their pre-feature state.
The figure's two output groups reflect this separation: the agent receives initial code, sequential specifications, and the container image, while reference code, hidden tests, and validation records remain private.

\input{03-benchmark-statistics}

%% file: fig-1-overview.tex
\begin{figure}[t]
  \centering
  \includegraphics[width=\columnwidth]{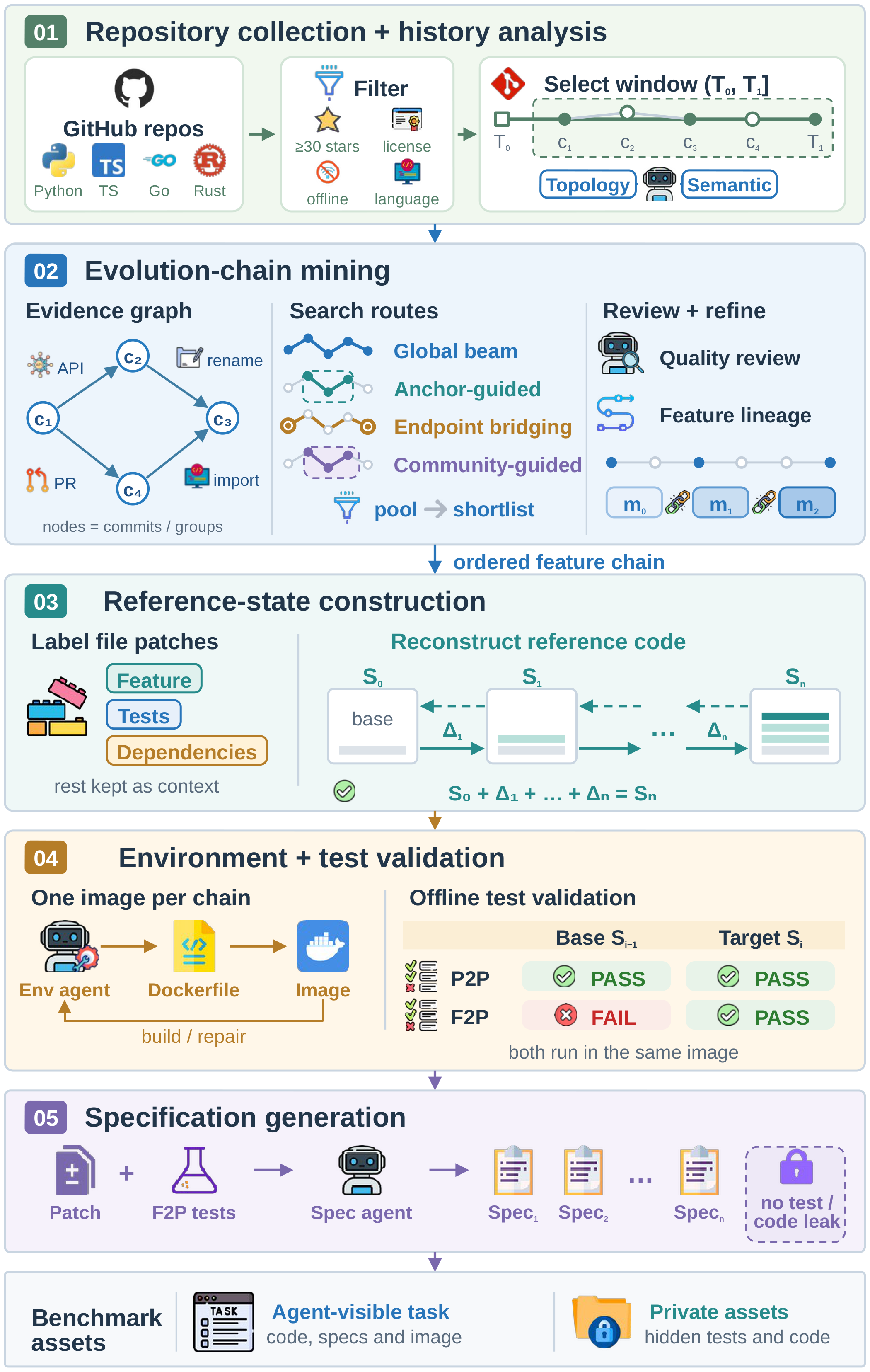}
  \caption{\textbf{Benchmark construction workflow.}}
  \Description{Five construction stages: repository collection and history
  analysis, evolution-chain mining, reference-state construction, environment and
  test validation, and specification generation. Step 1 filters repositories by
  stars, license, offline support, and language, selects a commit window, and has
  topology and semantic agents review its history. Within the mining region, a
  schematic graph links commits using API, rename, pull-request, and import evidence. It
  feeds global beam search, anchor-guided search,
  endpoint bridging, and community-guided search, followed by candidate
  retention and review and refinement. The outputs separate initial code,
  sequential specifications, and a shared image from reference states, stage
  suites, and validation evidence. The evaluated agent carries its own
  implementation forward.}
  \label{fig:construction-overview}
\end{figure}

%% file: 03-benchmark-statistics.tex
\subsection{Evaluation-Set Curation}
\label{sec:eval-set}

Multi-stage evaluation requires repeated model calls and sequential execution within each feature, with costs incurred for every evaluated agent.
Given these economic constraints, we select a high-quality subset of the constructed task pool as our evaluation set through three steps.

\textbf{Stratified sampling.}
To cover different languages and evolution lengths, we sample 100 candidate tasks, 25 per language, stratified by chain length.
This retains both short and longer feature-evolution chains in the candidate set.
A scan for test-used identifiers missing from the specifications identifies descriptions and contracts that need revision before execution.

\textbf{Task-defect exclusion.}
To identify task defects, we run candidate tasks with several models and inspect their execution trajectories.
We check these trajectories against the specifications, hidden tests, and reference implementations to distinguish task defects from agent errors.
This screening excludes seven tasks: five whose chain-level suites require unspecified behavior, one with a defective test suite, and one with a repository hook that intercepts the prompt.

\textbf{Manual check.}
To ensure that the selected tasks measure the intended feature-evolution setting, we manually check them for specification quality, alignment between interface contracts and hidden tests, and whether successive stages extend the same feature.
The resulting \benchname{} evaluation set contains \textbf{73 tasks with 263 stages from 53 repositories}.
All reported agent comparisons use this same set.

\subsection{Benchmark Characteristics}
\label{sec:snapshot}

\input{fig-task-stats}

\benchname{} supports evaluation of evolving-feature implementation through three complementary properties.
Figure~\ref{fig:task-stats} provides the supporting dataset characteristics.

\textbf{Historical same-feature dependencies.}
Later stages require agents to revisit earlier work, making inherited implementation decisions part of the task.
Every task connects successive stages through shared files or verified call dependencies.
As shown in Figure~\ref{fig:task-stats}d, a later stage shares, on average, 54.3\% of its changed files with earlier stages; the median across tasks is 59.1\%.
Tasks span 2 to 10 stages, with 3.60 on average, providing both short extensions and longer sequences of dependent development.

\textbf{Requirements-grounded verification.}
The benchmark combines explicit stage specifications with validated stage-level and whole-feature checks.
These tasks involve substantial changes across a repository: Figure~\ref{fig:task-stats}c shows median reference changes of 426 source lines and 416 test lines per stage, with whiskers at the 5th and 95th percentiles; Figure~\ref{fig:task-stats}d shows an average of 7.71 changed source files.
The median whole-feature change is 1,453 source lines.
Across implementation changes of this scale, paired-state validation and manual checks establish alignment between the executable evaluation and the behavior requested in the stage specifications.

\textbf{Diverse repository contexts.}
The tasks cover four languages and seven application domains, assigned per repository by hand, allowing the same evaluation process to examine different development settings.
Figure~\ref{fig:task-stats}b shows a spread across Python, Go, TypeScript, and Rust (21, 20, 18, and 14 tasks, respectively).
Figure~\ref{fig:task-stats}a includes AI agents and LLM tooling as well as cloud and infrastructure, with 17 and 14 tasks, respectively.
The shared task structure supports evaluation across varied code, interfaces, and application needs: each agent must extend one feature per task.

%% file: fig-task-stats.tex
\begin{figure}[t]
  \centering
  \includegraphics[width=\columnwidth]{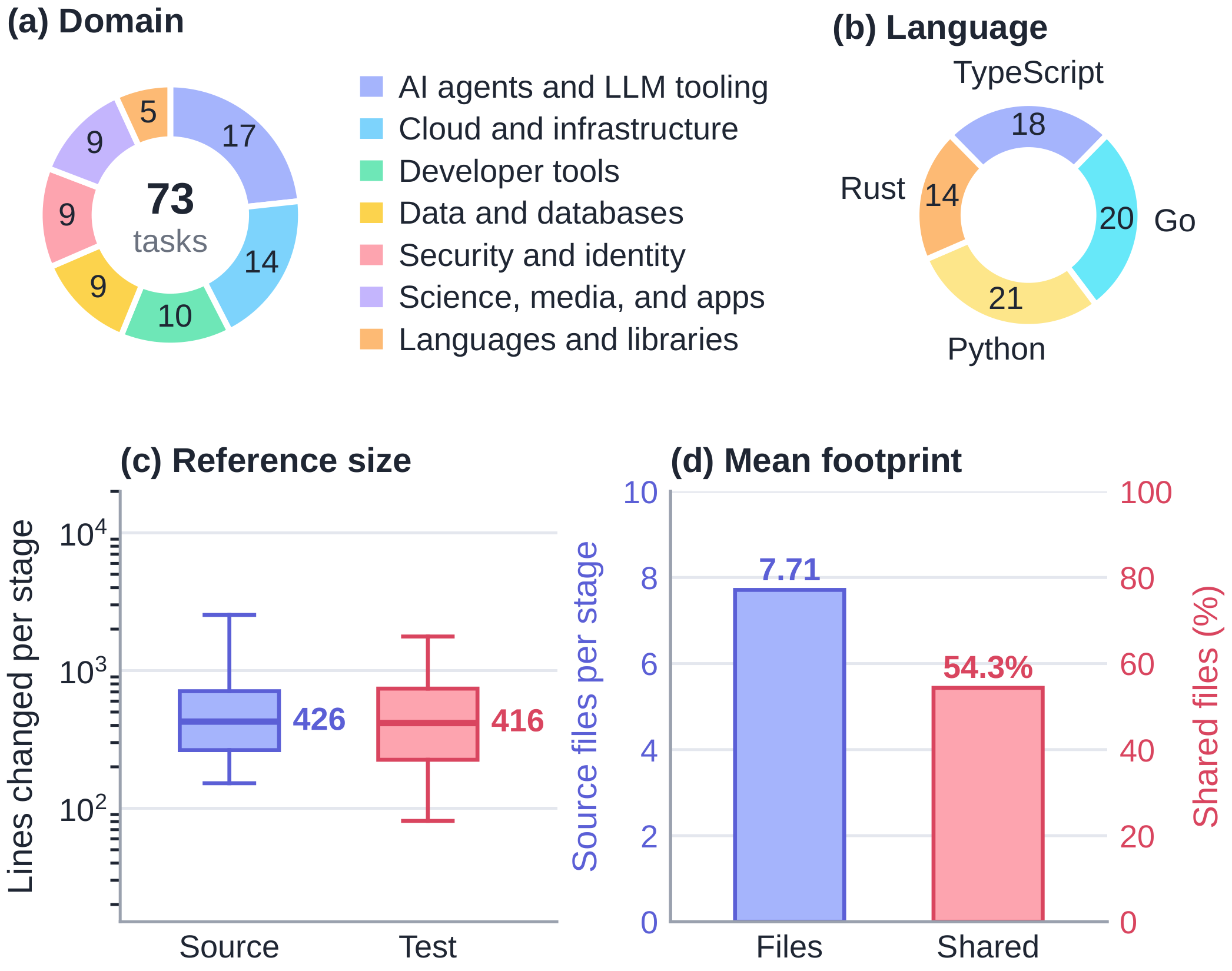}
  \caption{\textbf{Characteristics of \benchname{}.}}
  \Description{Two donut charts give tasks per domain and per language. A log-scale box plot gives source
  and test lines changed per stage, with medians 426 and 416, respectively. Two bars give 7.71 source files per stage and
  54.3 percent shared files.}
  \label{fig:task-stats}
\end{figure}

%% file: 04-study-design.tex
\section{Experimental Setup}
\label{sec:study}

We evaluate coding agents on \benchname{} using the L4 process in Section~\ref{sec:protocol}.
Our study addresses three complementary questions:
\begin{itemize}
  \item \textbf{RQ1: How reliably do agents complete evolving features?} Stage and chain gates measure implementation success, failure position, and the effect of task length.
  \item \textbf{RQ2: How do earlier errors affect later stages?} Conditional pass rates and inheritance labels distinguish defects carried forward from regressions introduced by later edits.
  \item \textbf{RQ3: How effectively do agents detect and verify their own errors?} Awareness labels, self-test outcomes, and specification checks examine whether agents recognize failures and obtain reliable evidence of correctness.
\end{itemize}

\subsection{Evaluated Agents}
\label{sec:agents}

To examine this setting across providers, we evaluate agents built on Claude Opus~5.5~\cite{anthropic2026opus55}, GPT-6 Astra~\cite{openai2026astra}, Hunyuan~4 Preview~\cite{tencent2026hunyuan}, GLM-5.3~\cite{zai2026glm53}, Grok~4.7~\cite{xai2026grok47}, Qwen~3.8~\cite{qwen2026qwen38}, Gemini~3.8 Flash~\cite{google2026gemini38}, Doubao~2.1 Pro~\cite{bytedance2026seed21}, Kimi~K3~\cite{moonshot2026kimik3}, and DeepSeek V4 Pro~\cite{deepseek2026v4pro}.
Hunyuan~4 Preview is the internal pre-release model evaluated in this study; the citation describes its model family.
The unit of comparison is the configured coding agent, including its model and harness.

\subsection{Trajectory Analysis}
\label{sec:labeling}

Gate outcomes locate failures, while trajectories show what the agent checked and whether it recognized an earlier error.
We use trajectory analysis to connect these observations for RQ2 and RQ3.
Process labels describe behavior visible in the specifications, trajectory, and agent code; diagnosis labels additionally use hidden-test evidence to explain the observed failure.

We use the following labels to describe verification behavior and diagnose failed stages.
\begin{itemize}\raggedright
  \item \textbf{Final check} records the strongest check the agent ran: relevant tests, unrelated tests, a manual behavior check, a build only, no check, or no activity. \textbf{Last test result} records whether the last test command passed, failed, collected no tests, was not run, or had an unclear result.
  \item \textbf{Criterion targeting} records whether the checks exercise all acceptance criteria, some criteria, only implementation details, or no behavior. It measures what the checks exercise; their expected outcomes may still be wrong.
  \item \textbf{Stop reason} records whether the agent ended after a successful check, ended without verification, dismissed a failure, ignored a failure, or did no work. \textbf{Dismissal evidence} records whether a claim that a failure was unrelated or pre-existing is supported by a baseline test, unverified, or contradicted by the trajectory.
  \item \textbf{Prior awareness} records whether an agent missed an earlier failure, noticed it without repairing it, or repaired it. It applies to stages following a failed stage.
  \item \textbf{Inheritance} records whether a failed later stage was caused wholly, partly, or not at all by an earlier defect, together with the stage that introduced that defect.
  \item \textbf{F2P cause: fine} records the specific reason for an F2P failure, such as a missing symbol, a wrong signature, incorrect logic, or a missed case. \textbf{F2P cause: grouped} combines these labels into interface, other build, and behavioral failures, while keeping timeouts, failures unrelated to the solver, and unknown causes separate.
\end{itemize}

\subsection{Implementation Details}
\label{sec:implementation}

\textbf{Task construction and review.}
The pipeline implements the five steps in Figure~\ref{fig:construction-overview} as agents with separate responsibilities for history review, feature-chain review, patch classification, environment preparation, test extraction, and specification writing.
Published task manifests record the models and artifacts used at each step.
We used GPT-5.5~\cite{openai2026gpt55} for Steps~1--3, DeepSeek V4.1 Flash~\cite{deepseek2026v41flash} for Step~4, and Claude Sonnet~5~\cite{anthropic2026models} for Step~5.
Acceptance of a task depends on executable reference-state validation and review of the task materials.

\textbf{Agent execution and comparability.}
Nine evaluated agents run in Claude Code,\footnote{Claude Code: \url{https://code.claude.com/docs/en/overview}.} while GPT-6 Astra uses the native Codex CLI.\footnote{Codex CLI: \url{https://github.com/openai/codex}.}
For GPT-6 Astra, Codex preserves the native tool-call representation and session-resumption mechanism, avoiding the additional adaptation required to run it through Claude Code.
Each harness resumes the same session at the next stage.
Each integration preserves its harness's handling of tool calls and conversation state, with context and reasoning controls specific to the provider.
Every agent is evaluated under the same task conditions: initial code, stage specifications, hidden acceptance checks, and a four-hour stage budget, with no external failure feedback and no rollback between stages.
Our principal comparisons follow each agent from its first stage to the whole feature and compare its behavior after passed and failed stages, keeping its configuration fixed.
Cross-agent scores reflect joint model--harness performance, with the two contributions coupled in these comparisons.
Tasks execute in offline containers with preinstalled dependencies, and hidden scoring runs separately from the agent's workspace.

\textbf{Trajectory annotation.}
Hunyuan~4 Preview, DeepSeek V4.1 Flash, and GLM-5.3 Flash~\cite{zai2026glm53} independently annotate the trajectories using a shared codebook, following the LLM-as-judge approach~\cite{zheng2023llmjudge}.
They assign process labels before accessing gate logs, hidden tests, and reference patches for diagnosis, and cite trajectory evidence for every label.
Reported labels use the majority decision; the first author adjudicates items on which all three judges disagree.
Analyses requiring a defect's origin use inherited failures whose originating stage can be identified.
To validate the labels, the first author independently annotates a stratified sample without seeing the model labels: passed stages, failed stages with no prior failure, and stages following a failure.
Agreement is measured before adjudication, using only applicable and resolved comparisons; dismissal evidence is compared when both judgments identify a dismissal.
An automatic classifier groups tool calls into exploration, editing, testing, building, and other actions.
Table~\ref{tab:label-validation} summarizes the blind review; Section~\ref{sec:annotation-results} reports further validation results.

\input{label-validation}

\textbf{Specification checks.}
For RQ3, we inspect cases in which a green self-test conflicts with the hidden F2P result.
Eligible stages begin at $m_0$ or after a passed stage and have checks labeled as targeting every acceptance criterion.
The first author examines the first failing expectation in a random sample, excluding chain runs used for blind label validation.
Each expectation is classified by its visible basis: stated in a specification, implied by the specifications, implied by the starting repository, or unsupported by those materials.

\subsection{Metrics and Statistics}
\label{sec:metrics}

We adopt the following metrics; table abbreviations appear in parentheses.
\begin{itemize}
  \item \textbf{Single-stage pass rate $m_0$:} the fraction of tasks whose first stage passes both F2P and P2P, starting from the initial state $S_0$.
  \item \textbf{Chain-gate pass rate (Gate):} the fraction of tasks whose chain gate passes on the final workspace.
  \item \textbf{Continuation rate (Cont.):} the stage pass rate at $m_1$ and later, conditioned on all earlier stages of that chain having passed.
  \item \textbf{Stage F2P rate (F2P):} the fraction of evaluated stages whose F2P suite passes.
  \item \textbf{Prefix depth (Depth):} the number of consecutive passed stages before the first failure, divided by chain length.
  \item \textbf{Survival at $m_k$:} the fraction of tasks with more than $k$ stages whose stages $m_0..m_k$ all pass.
  \item \textbf{Chain-level P2P failure (P2P$\downarrow$):} the fraction of tasks whose chain-level P2P suite fails.
  \item \textbf{In-chain regression rate:} among runs in which every stage passed F2P at completion, the fraction whose final workspace fails chain-level F2P.
  \item \textbf{Pure regression rate:} among runs whose final workspace passes chain-level F2P, the fraction that fail chain-level P2P.
\end{itemize}

For comparisons within \textbf{the same task and stage}, we split agents by the condition being compared and retain only task-stage groups containing both conditions.
We compute each condition's F2P pass rate within each group, then average these rates equally across groups.
The agents in each group face the same requested stage but retain their own code and conversation histories.

Confidence intervals for agent performance are 95\% bootstrap intervals~\cite{efron1979bootstrap} with 4,000 resamples over tasks in \benchname{}.

%% file: label-validation.tex
\begin{table}[!t]
  \caption{Blind validation of trajectory labels.}
  \label{tab:label-validation}
  \centering\small
  \setlength{\aboverulesep}{0pt}
  \setlength{\belowrulesep}{0pt}
  \begin{tabularx}{\columnwidth}{X ccc}
    \toprule
    \rowcolor{tablehead}
    \rowpadtop\rowpadbot\textbf{Label} & \textbf{Agree / $n$} & \textbf{\%} & \textbf{95\% CI} \\
    \midrule
    \multicolumn{4}{l}{\rowpadtop\textit{Process labels}} \\
    \rowcolor{tableband}
    Final check & 175/180 & 97.2 & 93.7--98.8 \\
    Last test result & 163/179 & 91.1 & 86.0--94.4 \\
    \rowcolor{tableband}
    Criterion targeting & 161/180 & 89.4 & 84.1--93.1 \\
    Stop reason & 165/180 & 91.7 & 86.7--94.9 \\
    \rowcolor{tableband}
    Dismissal evidence & 41/44 & 93.2 & 81.8--97.7 \\
    Prior awareness\rowpadbot & 56/60 & 93.3 & 84.1--97.4 \\
    \midrule
    \multicolumn{4}{l}{\rowpadtop\textit{Diagnosis labels}} \\
    \rowcolor{tableband}
    Inheritance & 90/90 & 100.0 & 95.9--100 \\
    F2P cause: fine & 88/96 & 91.7 & 84.4--95.7 \\
    \rowcolor{tableband}
    F2P cause: grouped\rowpadbot & 96/96 & 100.0 & 96.2--100 \\
    \bottomrule
  \end{tabularx}
\end{table}

%% file: 05-results.tex
\section{Experimental Results}
\label{sec:results}

We report the results on \benchname{} under the L4 evaluation process.
Tables~\ref{tab:delivery} and~\ref{tab:mechanism} report percentages with best values bold; Table~\ref{tab:delivery} is ordered by Gate.
The analysis first examines whole-feature completion (RQ1), then the effect of earlier errors (RQ2), and finally agents' ability to detect and verify those errors (RQ3).

\noindent\textbf{Trajectory-label validation.}
\label{sec:annotation-results}
Independent human annotations provide a check on the trajectory labels used to explain the evaluation outcomes.
Table~\ref{tab:label-validation} reports the blind review of 180 stages, with 60 from each sampling group defined in Section~\ref{sec:implementation}.
Agree/$n$ counts matches with the model majority among applicable comparisons; intervals are 95\% Wilson intervals~\cite{wilson1927probable}.
Agreement is 89.4\%--97.2\% for process labels and 91.7\%--100.0\% for diagnosis labels, over their applicable comparisons.
For the labels central to RQ2 and RQ3, the table reports 93.3\% agreement on prior awareness and 100.0\% on inheritance.
Across the three model judges, Fleiss' $\kappa$~\cite{fleiss1971kappa} is 0.94, 0.92, 0.91, 0.86, 0.84, and 0.80 for prior awareness, final check, stop reason, last test result, inheritance, and criterion targeting, respectively.
The first author adjudicated 132 label items across 121 stages on which the judges all differed.
The automatic tool classifier agrees with the trajectory labels on 94.8\% of stages for which both identify a final red/green test result.

\begin{table}[!ht]
  \caption{Performance on \benchname{} at L4.}
  \label{tab:delivery}
  \centering\small
  \setlength{\tabcolsep}{2.5pt}
  \setlength{\aboverulesep}{0pt}
  \setlength{\belowrulesep}{0pt}
  \begin{tabularx}{\columnwidth}{X cc ccc c}
    \toprule
    \rowcolor{tablehead}
    \rowpadtop\rowpadbot & \multicolumn{2}{c}{\textbf{Chain}} & \multicolumn{3}{c}{\textbf{Stage level}} & \\
    \cmidrulefill{tablehead}
    \cmidrule(lr){2-3}\cmidrule(lr){4-6}
    \rowcolor{tablehead}
    \rowpadtop\rowpadbot\textbf{Model} & $m_0$ & Gate & Cont. & F2P & Depth & P2P$\downarrow$ \\
    \midrule
    \rowpadtop \modellogo{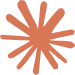}Claude Opus 5.5   & \textbf{83.6} & \textbf{49.3} & \textbf{79.2} & \textbf{68.4} & \textbf{64.9} & \textbf{5.5} \\
    \rowcolor{tableband}
    \modellogo{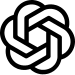}GPT-6 Astra       & 67.1 & 37.0 & 76.1 & 55.5 & 50.4 & 20.5 \\
    \modellogo{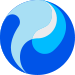}Hunyuan 4 Preview & 60.3 & 34.2 & 75.0 & 49.4 & 44.1 & 8.2 \\
    \rowcolor{tableband}
    \modellogo{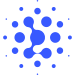}GLM-5.3           & 64.4 & 31.5 & 72.0 & 56.3 & 45.8 & 19.2 \\
    \modellogo{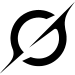}Grok 4.7          & 60.3 & 26.0 & 70.9 & 50.6 & 40.9 & 11.0 \\
    \rowcolor{tableband}
    \modellogo{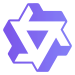}Qwen 3.8          & 61.6 & 24.7 & 65.4 & 49.0 & 39.5 & 6.8 \\
    \modellogo{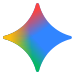}Gemini 3.8 Flash  & 61.6 & 23.3 & 66.2 & 49.4 & 42.0 & \textbf{5.5} \\
    \rowcolor{tableband}
    \modellogo{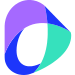}Doubao 2.1 Pro    & 60.3 & 21.9 & 60.0 & 43.0 & 36.1 & 8.2 \\
    \modellogo{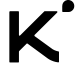}Kimi K3           & 56.2 & 19.2 & 64.5 & 42.2 & 35.7 & 15.1 \\
    \rowcolor{tableband}
    \modellogo{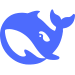}DeepSeek V4 Pro\rowpadbot & 52.1 & 17.8 & 61.5 & 40.3 & 31.6 & 11.0 \\
    \midrule
    \rowcolor{tablesummary}
    \rowpadtop\rowpadbot\textbf{Pooled}   & 62.7 & 28.5 & 70.1 & 50.4 & 43.1 & 11.1 \\
    \bottomrule
  \end{tabularx}
\end{table}

\subsection{RQ1: Completing Evolving Features}
\label{sec:delivery}

To determine whether success on one request carries over to an evolving feature, we compare first-stage and chain-gate performance in Table~\ref{tab:delivery}.
Figures~\ref{fig:survival} and~\ref{fig:outcomes} then relate this gap to stage position and task length.
Figure~\ref{fig:survival} pools agents, using the 57 and 34 tasks containing $m_2$ and $m_3$, respectively, at those stages.
Figure~\ref{fig:outcomes}b compares first-stage (hollow) and chain-gate (filled) scores with task-bootstrap intervals; $\Delta$ is their percentage-point difference.


\input{fig-survival}
\input{fig-chain-outcomes}

\textbf{Finding 1: The chain-gate pass rate falls to less than half of the first-stage pass rate.}
The first stage starts from $S_0$ and is equivalent to a standard single-stage task.
Pooled over agents, success drops from 62.7\% at the first stage to 28.5\% at the chain gate (Table~\ref{tab:delivery}).
Every agent loses 26.0--38.4 percentage points (Figure~\ref{fig:outcomes}b), and even Claude Opus 5.5, which leads every metric, fails more than half of the tasks.
Rankings stay broadly similar (Spearman $\rho{=}0.87$ between $m_0$ and the chain gate), but first-stage scores overstate how often agents complete a feature.

\textbf{Finding 2: Checking only the last stage overstates success.}
The last-stage gate checks the final extension using its stage-level F2P and P2P suites, whereas the chain gate checks the complete feature on the final workspace.
Of the 273 runs that pass the last-stage gate, 69 (\textbf{25.3\%}) fail the chain gate.
The separate whole-feature check in Figure~\ref{fig:protocol} detects this mismatch: passing the last extension does not establish that the complete feature works on the final code.
Recovery is possible but uncommon: among the 525 runs with at least one failed stage, only 14 (\textbf{2.7\%}) eventually pass the chain gate, the segment marked ``Rec.'' in Figure~\ref{fig:outcomes}a.
``Gate fails'' marks runs passing every stage gate but failing the chain gate.

\textbf{Finding 3: Most runs break in the first two stages, and long tasks rarely succeed.}
Figure~\ref{fig:outcomes}a shows where runs that never recover first fail.
In 36.2\% of all runs, the agent already fails $m_0$; in another 20.7\%, it first fails $m_1$, the first stage that builds on its own code.
Only 13.2\% of runs first fail at a later stage.
The declining survival curves in Figure~\ref{fig:survival}a show how these early failures reduce the share of chains that remain correct as requirements accumulate.
Success also falls steeply with task length (Figure~\ref{fig:survival}b): the chain gate passes in 44.4\% of runs on two-stage tasks but in only \textbf{1.5\%} on tasks with five or more stages; only two runs pass, both by Grok~4.7.
Completing these longer features remains difficult at L4.

\rqanswer{RQ1}{At L4, agents complete a whole evolving feature far less often than they pass its first stage: pooled success falls from 62.7\% to 28.5\%. Most runs break at $m_0$ or $m_1$, runs with a failed stage rarely recover fully, and tasks with five or more stages are almost never completed.}

\subsection{RQ2: The Effect of Earlier Errors}
\label{sec:loss}

\input{fig-inheritance}

To explain failures beyond the first stage, we examine whether later stages build on correct behavior or inherit unresolved defects.
Figure~\ref{fig:inheritance} compares next-stage success after passed and failed stages, and Table~\ref{tab:mechanism} reports the trajectory evidence.
The agents revisit earlier code to a similar extent as the reference implementations: 52.8\%--57.7\% of their later-stage changed files were changed earlier, consistent with the dependence shown in Figure~\ref{fig:task-stats}d.
This establishes the setting in which an earlier implementation error can affect a later request.

\input{mechanism}

\textbf{Finding 4: Inherited defects are common in failed later stages.}
Next-stage F2P success falls from \textbf{70.2\%} after a passed stage to \textbf{22.6\%} after a failed stage (Figure~\ref{fig:inheritance}).
The gap appears for every agent.
Using the comparison within each task and stage defined in Section~\ref{sec:metrics}, we retain 109 task-stage groups containing both prior-stage outcomes.
The gap remains when task-stage groups receive equal weight: the averaged within-group F2P pass rates are 72.6\% after a passed stage and 29.0\% after a failed stage.
Once a run fails a stage, on average 85.1\% of its remaining stages fail as well.
The inheritance column of Table~\ref{tab:mechanism} connects this performance gap to concrete defects: 72.9\% of failed later stages involve errors left in earlier stages, with those defects accounting for the failure wholly in 47.1\% and partly in 25.9\%.

We also examine how stage position and context relate to success in later stages.
Among runs in which all earlier stages passed, the stage-gate pass rates at $m_1$, $m_2$, and $m_{3+}$ range from 66.6\% to 77.7\%, with no downward trend.
For later stages with available context measurements, the F2P gap between stages with and without detected context compaction is 6.2 percentage points for stages with fewer than 40 tool calls and 5.2 points for those with 40--99 tool calls.
Among later stages with fully passed prefixes, F2P success is 75.0\% when the cumulative specifications contain 1,500--2,999 words and 68.6\% when they contain at least 8,000 words, a difference of 6.4 points.
These gaps are smaller than the 47.6-point gap associated with the previous stage's outcome.

\textbf{Finding 5: Regression is uncommon after stages pass their tests.}
Regression, in which later edits break behavior that already worked, is the other way a chain can fail. We measure it in two ways.
\textbf{In-chain regression}: every stage passed its own F2P suite, yet the final workspace fails the chain-level F2P suite; it occurs in 7.3\% of the runs in which every stage passed F2P.
\textbf{Pure regression}: the final workspace passes chain-level F2P but fails chain-level P2P; it occurs in 3.3\% of the runs whose final workspace passes chain-level F2P.
Inherited defects fail at least one later stage in 52.9\% of feature runs, consistent with the inheritance evidence in Table~\ref{tab:mechanism}.
Runs that establish the required behavior usually retain it through the final check.
The higher P2P failure rates in Table~\ref{tab:delivery} (5.5\%--20.5\%) also include runs whose new feature behavior fails; they answer a broader question than pure regression.

\rqanswer{RQ2}{Earlier defects frequently affect later development: next-stage F2P success falls from 70.2\% after a passed stage to 22.6\% after a failed one, and trajectory analysis identifies inherited defects in 72.9\% of failed later stages. Regression after successful implementation is less common under the conditional measures examined here.}

\subsection{RQ3: Detecting and Verifying Errors}
\label{sec:self-verification}

To understand why agents continue from defective code, we examine whether they notice earlier failures and whether their own checks provide reliable evidence of correctness.
Table~\ref{tab:mechanism} connects awareness and verification behavior to hidden-test outcomes; lower inheritance and missed-prior-failure rates are marked as better.
Figure~\ref{fig:selftest} pools later stages across agents, showing hidden F2P outcomes conditional on the final self-test and previous-stage results.

\textbf{Finding 6: Agents rarely notice their own earlier errors.}
For stages following a failure, the prior-awareness label shows that the agent never notices the earlier error in \textbf{94.9\%} of cases (90.2\%--99.0\% per agent, Table~\ref{tab:mechanism}); the blind review agrees with these labels on 56 of 60 sampled stages (Table~\ref{tab:label-validation}).
Of the 50 stages in which the agent does notice the error, it repairs the error in 34.
These repair cases show that correction is possible once an error is recognized; the much larger missed-failure group places the more frequent breakdown at detection.

Tool use changes little after a failed stage.
For the nine agents on Claude Code, the median number of tool calls is 4\%--18\% higher after a failed stage than after a passed one, and the share of calls spent on testing barely changes (15.0\% versus 15.3\% for Claude Opus 5.5).

\input{fig-selftest}

\textbf{Finding 7: A green self-test barely predicts the hidden F2P result, and a red one is usually dismissed.}
Agents stop when their own tests pass, but a passing self-test says little about the hidden F2P suite.
Overall, 73.8\% of stages end on a green self-test, yet only \textbf{52.7\%} of these pass F2P, barely above the 50.4\% rate over all stages; stages that end red still pass in \textbf{45.2\%} (Table~\ref{tab:mechanism}).
Green/Red in Table~\ref{tab:mechanism} are conditional F2P rates; Ends green is the share ending green. Red estimates use fewer than 20 stages for Grok 4.7 and Gemini 3.8 Flash.
Using the same comparison procedure, the mean within-group F2P pass rates are 52.9\% for green final self-tests and 47.9\% for red final self-tests across 161 task-stage groups containing both outcomes.
As Figure~\ref{fig:selftest} shows, the self-test is least reliable right after a failed stage: of the 713 stages that still end on a green self-test, only 24.5\% pass F2P. The share of false greens, green self-tests whose stage fails F2P, rises from 28.7\% after a passed stage to 75.5\% after a failed one.

A red self-test, when it occurs, is usually dismissed.
Of the 367 stages that end red, the agent states that the failure is unrelated or pre-existing in 321 (87.5\%).
Only 23 of these 321 (7.2\%) first ran the same test on the stage's starting code to confirm that it already failed there.
The final self-test result is therefore a weak signal of correctness, and a red result often leads to dismissal.

\textbf{Finding 8: Self-tests that target every criterion still miss what the specification asks.}
The criterion-targeting label marks 68.4\% of stages, and 30.4\%--84.8\% per agent, as having checks that exercise all acceptance criteria (Table~\ref{tab:mechanism}).
Targeting all criteria offers limited assurance: these stages pass F2P in 54.9\% of cases, against 42.1\% when the checks exercise only some criteria.
Using the same comparison procedure, the mean within-group F2P pass rates are 50.4\% for all criteria and 38.5\% for some criteria across 190 task-stage groups containing both categories.
Of the stages that target all criteria, 45\% still fail F2P, and 62\% of these failures follow an earlier failed stage that the agent did not notice.

The mismatch also appears at the first stage and after passed stages.
Criterion coverage and the correctness of test expectations are distinct: a check can target the requested behavior while encoding an incorrect expected result.
In 239 stages, the checks exercise all criteria, the stage begins at $m_0$ or after a passed stage, the self-test ends green, and F2P still fails; most are wrong logic or a misread specification (54.4\%) or a missed case or partial implementation (24.3\%).
In the 60-stage specification check described in Section~\ref{sec:implementation}, 34 of the first failing expectations are stated explicitly in a current or earlier specification, 12 follow from these specifications, and 14 follow from the starting repository.

The same gap appears before any test runs.
Of the 1,304 stages that fail F2P, 236 (18.1\%) fail because the hidden suite does not compile; 228 of these are caused by the agent, and interface mismatches cause 61.4\% of those 228: missing symbols or exports, wrong parameter or return types, and duplicate declarations.
Of the missing names, 44.1\% appear in the current stage's specification and 30.5\% in an earlier stage's.
Together with the criterion-targeting and green-test columns of Table~\ref{tab:mechanism}, these cases show why checking an agent's interpretation of a requirement may still miss the behavior stated in the task.
\benchname{} makes this discrepancy observable through explicit contracts and separate acceptance checks.

\rqanswer{RQ3}{Agents leave earlier errors unnoticed in 94.9\% of stages following a failure. Their self-tests provide limited assurance: only 52.7\% of stages ending on a green self-test pass F2P, and this falls to 24.5\% after a failed stage. Even checks labeled as targeting every criterion can miss specified behavior, while final red results are usually dismissed without a baseline check.}

%% file: fig-survival.tex
\begin{figure}[t]
  \centering
  \includegraphics[width=\columnwidth]{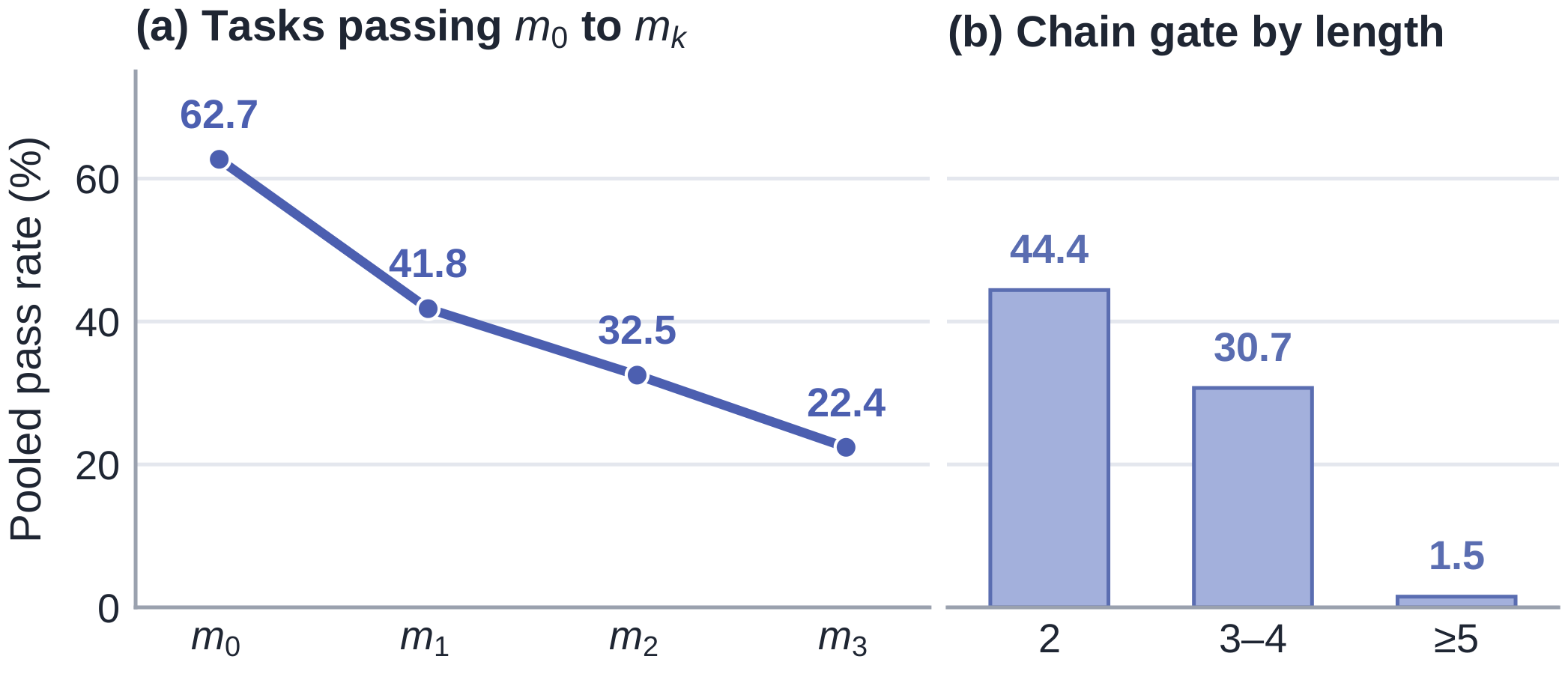}
  \caption{\textbf{Performance across stages and task lengths.}}
  \label{fig:survival}
\end{figure}

%% file: fig-chain-outcomes.tex
\begin{figure*}[t]
  \centering
  \includegraphics[width=\textwidth]{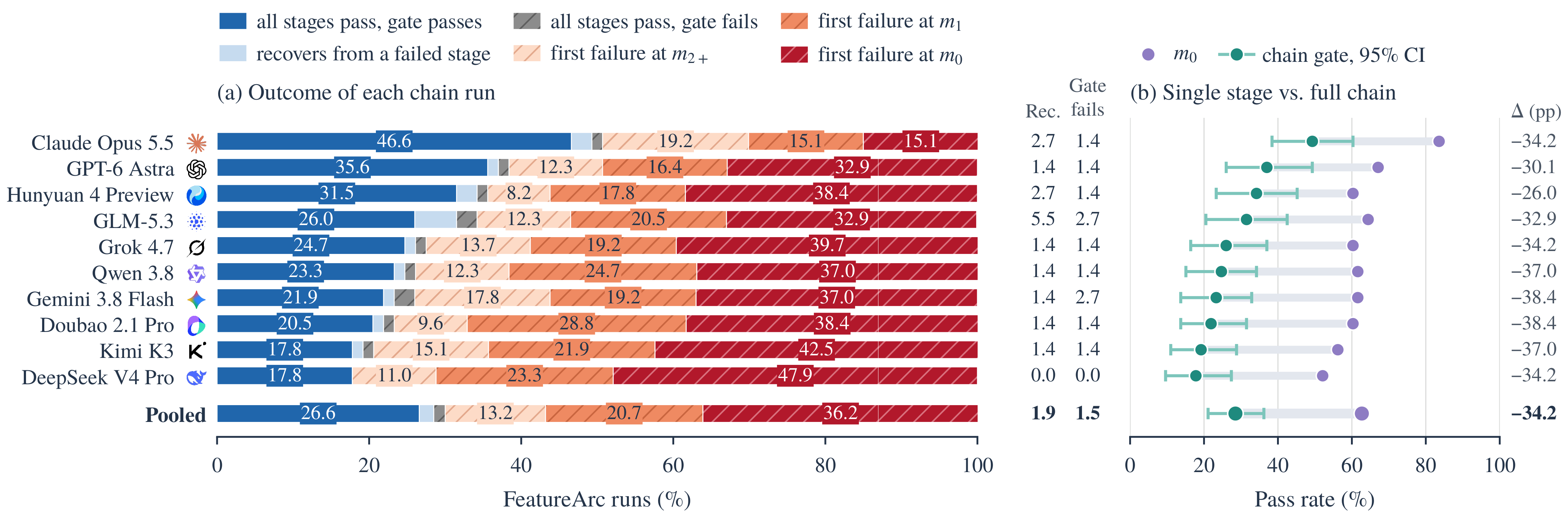}
  \caption{\textbf{Where tasks break.}}
  \label{fig:outcomes}
\end{figure*}

%% file: fig-inheritance.tex
\begin{figure}[t]
  \centering
  \includegraphics[width=\columnwidth]{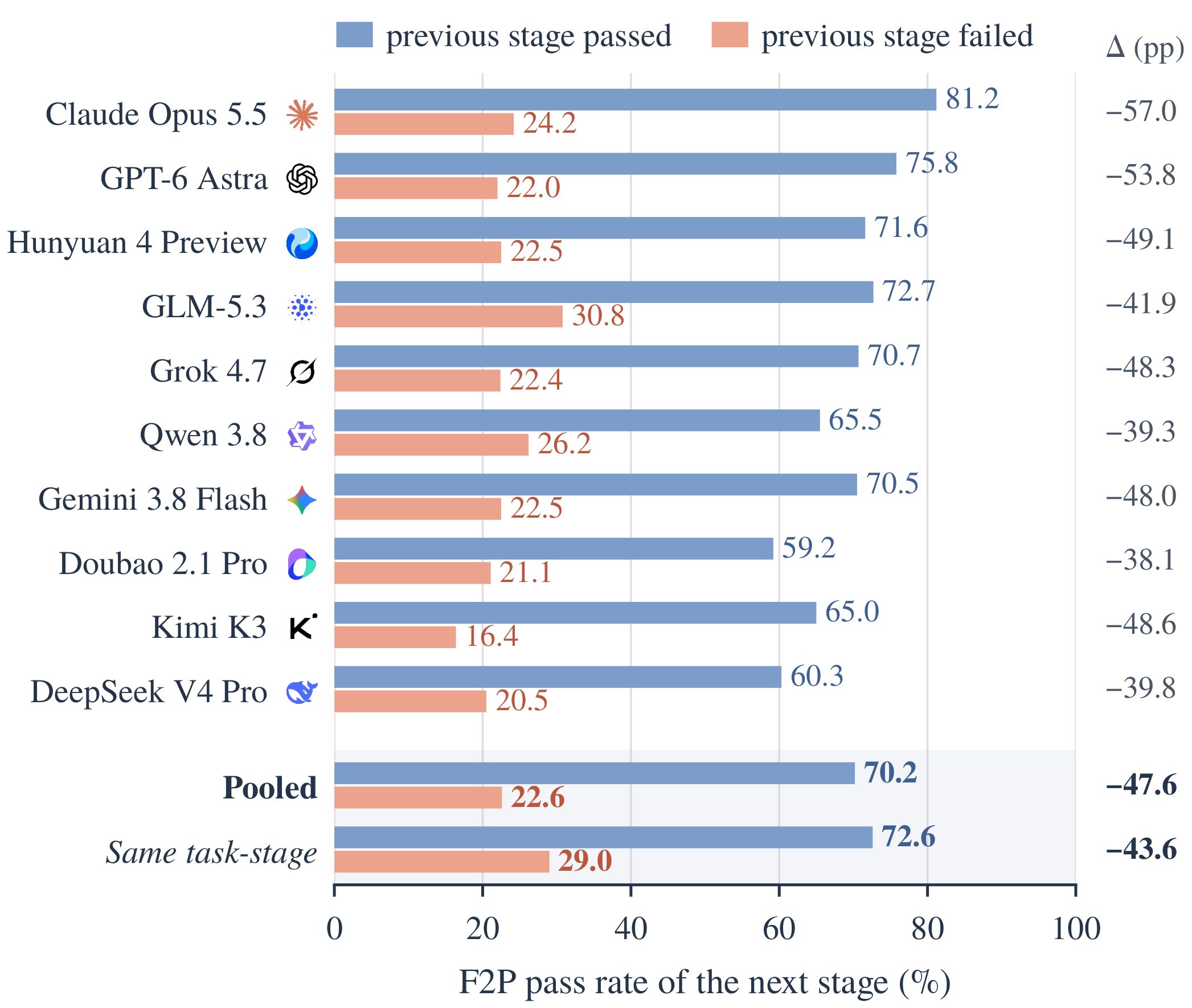}
  \caption{\textbf{Next-stage F2P success by the previous stage's outcome.}}
  \label{fig:inheritance}
\end{figure}

%% file: mechanism.tex
\begin{table}[t]
  \caption{Error inheritance and self-verification on FeatureArc.}
  \label{tab:mechanism}
  \centering\scriptsize
  \setlength{\tabcolsep}{3.9pt}
  \setlength{\aboverulesep}{0pt}
  \setlength{\belowrulesep}{0pt}
  \begin{tabular}{lcccccc}
    \toprule
    \rowcolor{tablehead}
    \rowpadtop\rowpadbot & \multicolumn{4}{c}{\textbf{Self-test}} & \multicolumn{2}{c}{\textbf{Trajectory labels}} \\
    \cmidrulefill{tablehead}
    \cmidrule(lr){2-5}\cmidrule(lr){6-7}
    \rowcolor{tablehead}
    \rowpadtop & \multicolumn{2}{c}{F2P after} & Ends & Criterion & & Prior \\
    \rowcolor{tablehead}
    \rowpadbot\textbf{Model} & Green & Red & green & targeting & Inheritance & awareness \\
    \midrule
    \rowpadtop \modellogo{opus55}Claude Opus 5.5 & \textbf{68.6} & 68.1 & 70.3 & 76.4 & 73.8 & 95.2 \\
    \rowcolor{tableband}
    \modellogo{gpt6}GPT-6 Astra & 54.5 & 46.4 & 81.0 & 82.9 & 74.8 & 95.6 \\
    \modellogo{hy4}Hunyuan 4 Preview & 52.6 & 41.3 & 65.8 & 70.3 & 73.7 & \textbf{90.2} \\
    \rowcolor{tableband}
    \modellogo{glm53}GLM-5.3 & 59.0 & 51.0 & 77.9 & \textbf{84.8} & \textbf{68.2} & 94.5 \\
    \modellogo{grok47}Grok 4.7 & 57.1 & 38.9 & 70.0 & 62.7 & 77.6 & 99.0 \\
    \rowcolor{tableband}
    \modellogo{qwen0902}Qwen 3.8 & 51.0 & 42.1 & 77.6 & 78.3 & 71.9 & 93.2 \\
    \modellogo{gemini38}Gemini 3.8 Flash & 49.6 & 20.0 & 97.3 & 77.2 & 71.9 & 97.1 \\
    \rowcolor{tableband}
    \modellogo{doubao21}Doubao 2.1 Pro & 47.2 & 34.6 & 68.4 & 70.0 & 73.1 & 90.4 \\
    \modellogo{kimi}Kimi K3 & 44.9 & 48.0 & 67.7 & 50.6 & 69.7 & 98.2 \\
    \rowcolor{tableband}
    \rowpadbot \modellogo{dsv4pro}DeepSeek V4 Pro & 41.5 & 32.4 & 62.4 & 30.4 & 75.0 & 96.6 \\
    \midrule
    \rowcolor{tablesummary}
    \rowpadtop\rowpadbot\textbf{Pooled} & 52.7 & 45.2 & 73.8 & 68.4 & 72.9 & 94.9 \\
    \bottomrule
  \end{tabular}
\end{table}

%% file: fig-selftest.tex
\begin{figure}[t]
  \centering
  \includegraphics[width=\columnwidth]{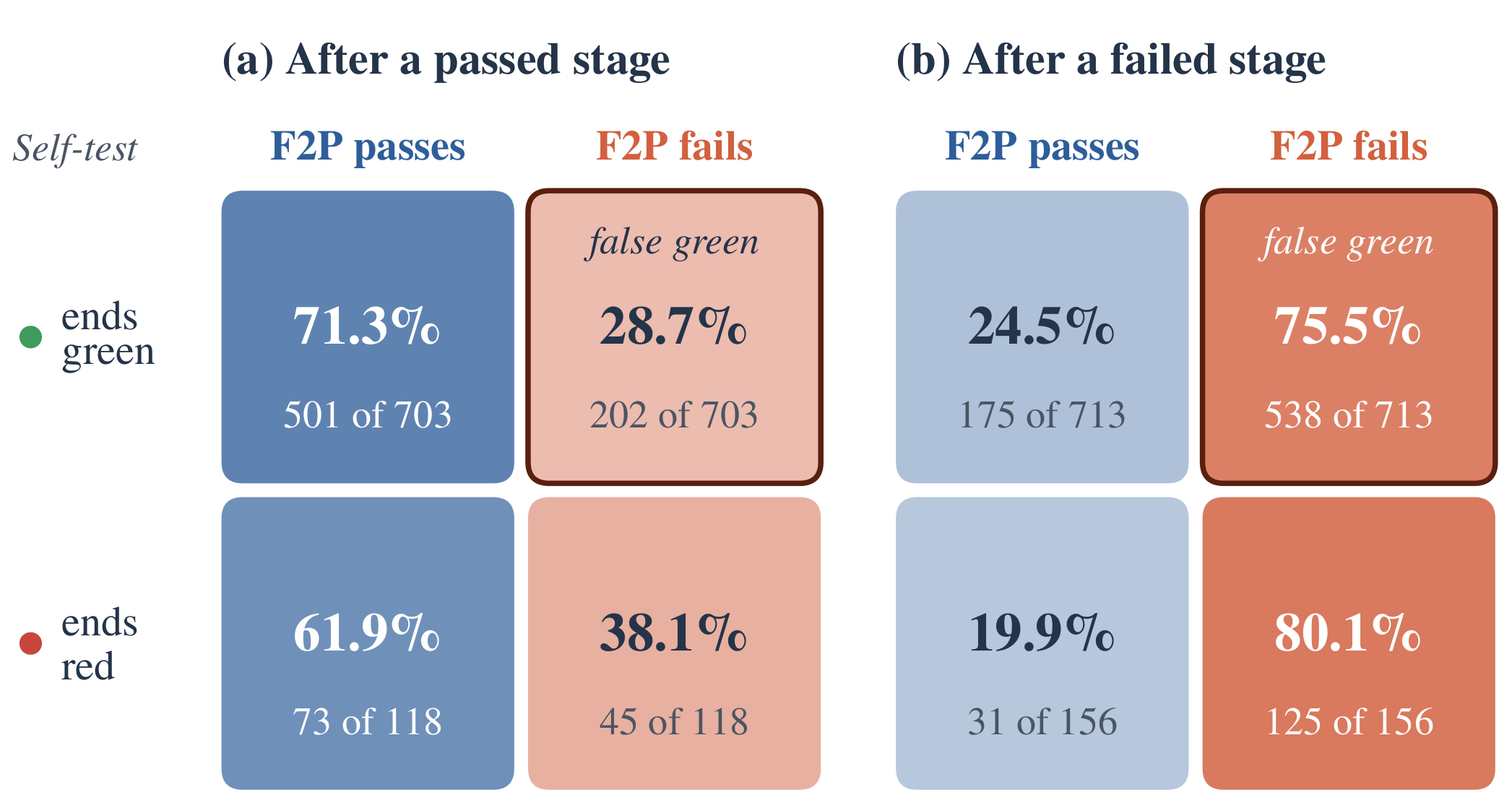}
  \caption{\textbf{Self-tests and hidden F2P outcomes.}}
  \label{fig:selftest}
\end{figure}

%% file: 06-lessons.tex
\section{Discussion}
\label{sec:tencent}

\subsection{Industry Application at Tencent}

\textbf{Connecting observed usage to internal evaluation.}
In an internal corpus of 8,881 IDE episodes at Tencent, 3,797 of the 8,728 episodes with valid user turns (43.5\%) contained follow-up interaction.
Model-based annotations of the first follow-up identified constraint additions (30.1\%), requirement clarifications (19.7\%), and functionality extensions (17.9\%).
The observed extensions motivate evaluating continued feature development.
To reduce user checking between extensions, agents must detect defects in their own earlier code.
Tencent's existing L1--L3 evaluations supply failure feedback, leaving this unaided detection ability untested.
We therefore use our pipeline and \benchname{} to assess Hunyuan~4 Preview alongside external agents at L4: users supply successive requirements, while earlier failures are not reported to the agent.

\textbf{Diagnosing development needs beyond the first-stage score.}
Table~\ref{tab:delivery} shows that Hunyuan~4 Preview passes 60.3\% of first stages but completes 34.2\% of whole features (95\% interval 23.3\%--45.2\%).
Its 75.0\% continuation rate indicates that it can often extend a feature when its earlier stages are correct.
Figure~\ref{fig:inheritance} reveals a different situation after an error: next-stage F2P success falls from 71.6\% after a passed stage to 22.5\% after a failed one.
The prior-awareness evidence in Table~\ref{tab:mechanism} helps explain this gap: Hunyuan leaves 90.2\% of earlier failures unnoticed.
This is the lowest missed-failure rate among the evaluated agents; most earlier errors still persist unrecognized during subsequent feature development.

\textbf{Turning evaluation evidence into development priorities.}
The pipeline links unsuccessful runs to specifications, code changes, and checks, helping the Hunyuan team distinguish failures to implement new behavior from failures to recognize inherited defects.
For example, Hunyuan attributes a final failing self-test to unrelated or pre-existing causes in 44 of 46 stages, but checks that explanation against the starting code in only 5 of those 44 cases.
This evidence gives the Hunyuan team concrete failure cases for improving checks of earlier requirements and verification of failure explanations.
These capabilities are necessary for supporting continued feature development when users leave intermediate validation to the agent.

\input{07-threats}

%% file: 07-threats.tex
\subsection{Threats to Validity}
\label{sec:threats}

\benchname{} may not cover the full range of workflows encountered in real-world feature development.
To broaden coverage, it includes four widely used programming languages---Python, TypeScript, Rust, and Go---and seven application domains, with tasks stratified by language and chain length (Figure~\ref{fig:task-stats}).
This breadth, together with tasks mined from repository histories, helps the benchmark reflect agents' performance on evolving requirements across practical development settings.

%% file: 08-related-work.tex
\section{Related Work}
\label{sec:related}

\textbf{Feature and interactive benchmarks.}
Single-request benchmarks evaluate feature implementation on fixed codebases~\cite{li2025feabench,chen2025featbench,zhou2026featurebench}.
Multi-stage feature~\cite{codecrafters2026ccbench,shastry2026swesteps,orlanski2026slopcodebench,nukrai2026evoweb} and sequential-development benchmarks~\cite{deng2026swemilestone,lam2026swechain,jin2026chainswe,le2026sweevo,xu2026roadmapbench,li2026loopsbench} extend evaluation across development stages, with different stage sources, feature scope, and feedback.
EvoCode-Bench allows four attempts that stop at the first failure and compares with reference-code starts~\cite{shen2026evocode}.
Interactive benchmarks add responsive users~\cite{raghavendra2026sweinteract,wu2026swetogether}, incremental requirements~\cite{zhan2026sreval,yan2026specemerges}, or clarification~\cite{vijayvargiya2026ambigswe}.

\textbf{Task construction.}
Environment-synthesis methods prepare executable repositories~\cite{hu2025repo2run,bouzenia2025executionagent,guo2026menvagent,eliseeva2025envbench}, while automated pipelines collect or synthesize individual tasks~\cite{badertdinov2025swerebench,badertdinov2026swerebenchv2,zhang2025swebenchlive,wang2025swebenchpp,guo2026swefactory,yang2025swesmith,jain2025r2egym,pan2025swegym,qi2026change2task}.
DeepCommit~\cite{deng2026swemilestone} and WEAVE~\cite{li2026weave} recover dependencies from repository histories; WEAVE uses them to award partial credit within a task.

\textbf{Self-verification.}
Execution feedback supports self-repair~\cite{chen2023selfdebug,shinn2023reflexion}, but its effectiveness depends on feedback quality~\cite{olausson2024selfrepair}.
Weak tests can overestimate correctness~\cite{liu2023evalplus}, and generating checks that expose defects remains difficult~\cite{mundler2025swtbench}.

\benchname{} differs by combining historical same-feature dependencies with continued evaluation on the agent's own code without intermediate failure feedback (Table~\ref{tab:benchmark-comparison}).
Validated stage and whole-feature checks measure correctness across extensions, while trajectory analysis examines the persistence and detection of inherited errors.

%% file: 09-conclusion.tex
\section{Conclusion}
\label{sec:conclusion}

We presented \benchname{} together with a pipeline that constructs its executable feature-evolution tasks from repository histories.
The benchmark combines historical same-feature dependencies, requirements-grounded verification, and diverse repository contexts to evaluate agents that extend their own code without external acceptance feedback.
Our study reveals a substantial gap between first-stage implementation and whole-feature completion: pooled success falls from 62.7\% to 28.5\%.
Inherited defects are common in failed later stages, and agents rarely notice these errors or expose them through self-tests.
At Tencent, the pipeline extends internal evaluation to L4 and provides evidence about Hunyuan's ability to recognize and recover from errors during feature development.
\benchname{} supports further study of sustained feature correctness as requirements evolve, using dependent stages and executable checks.